\documentclass[authoryear,times,5p]{elsarticle}

\usepackage{graphicx}

\usepackage{amssymb}
\usepackage{amsmath}

\newcommand \ie{\textit{i.e.}, }
\newcommand \eg{\textit{e.g.}, }

\journal{Icarus (with some subsequent modification differing from the original submission)}

\begin{document}

\begin{frontmatter}



\title{The Fermi Paradox, Self-Replicating Probes,\\and the Interstellar Transportation Bandwidth}


\author{~Keith B. Wiley}
\ead{kwiley@keithwiley.com}

\address{Univ. of Washington, Dept. of Astronomy, Box 351580, Seattle WA 98195, USA}

\begin{abstract}
It has been widely acknowledged that self-replicating space-probes (SRPs) could explore the galaxy very quickly relative to the age of the galaxy.  An obvious implication is that SRPs produced by extraterrestrial civilizations should have arrived in our solar system millions of years ago, and furthermore, that new probes from an ever-arising supply of civilizations ought to be arriving on a constant basis.  The lack of observations of such probes underlies a frequently cited variation of the Fermi Paradox.  We believe that a predilection for ETI-optimistic theories has deterred consideration of incompatible theories.  Notably, SRPs have virtually disappeared from the literature.  In this paper, we consider the most common arguments against SRPs and find those arguments lacking.  By extension, we find recent models of galactic exploration which explicitly exclude SRPs to be unfairly handicapped and unlikely to represent natural scenarios.

We also consider several other models that seek to explain the Fermi Paradox, most notably percolation theory and two societal-collapse theories.  In the former case, we find that it imposes unnatural assumptions which likely render it unrealistic.  In the latter case, we present a new theory of \textit{interstellar transportation bandwidth} which calls into question the validity of societal-collapse theories.

Finally, we offer our thoughts on how to design future SETI programs which take the conclusions of this paper into account to maximize the chance of detection.
\end{abstract}


\end{frontmatter}


\section{Introduction}
\label{label_section_Introduction}

The Fermi Paradox is the apparent absence of evidence for extraterrestrial intelligence (ETI) in our galaxy, or at least in our local neighborhood, despite calculations suggesting that galactic colonization should be feasible within the age of the galaxy.  Numerous theories, aside from an extraordinary rarity of ETI, have been offered to explain the lack of observables.  The willingness toward ETI-optimism is understandable not only because the alternative is undesirable, but because it is fairly unactionable; the null conclusion prescribes no future experiments.  Nevertheless, we should not be overly exclusive of incompatible yet reasonable theories of galactic exploration.  In section \ref{label_section_VonNeumannSelfReplicatingProbes} we consider that one such theory, the use of robotic self-replicating space-probes (SRPs) for exploration, has all but disappeared from the literature, leaving a gap in our investigation of the Fermi Paradox.

In section \ref{label_section_PercolationTheory} we consider a previous explanation for the Fermi Paradox based on a percolation model.

In section \ref{label_section_InterstellarSocietalCollapse} we evaluate theories of interstellar societal-collapse which attempt to resolve the Fermi Paradox.  In response, we present a new theory, the \textit{interstellar transportation bandwidth} (ITB), which we believe provides fresh insight into the feasibility of such models.  We also offer a refined version of the Drake equation which incorporates the ITB.

In section \ref{label_section_BonusStimulation} we consider an analysis of the Fermi Paradox based on mutual benefit following contact.

Finally, in section \ref{label_section_ETIMayStillExistinourGalaxy} we cite a previous theory which we believe best resolves the Fermi Paradox without excluding intragalactic ETI, and in section \ref{label_section_SETI} we discuss how SETI might benefit from the conclusions of this paper.

\section{Von Neumann Self-Replicating Probes}
\label{label_section_VonNeumannSelfReplicatingProbes}

The Fermi Paradox is greatly exacerbated by the proposition that galactic colonization might proceed by purposeful exploration and expansion as opposed to mere population diffusion.  Rapid exploration would logically benefit from robotic probes.  Such a proposition closely mirrors our own space exploration efforts to date.

In this paper we refer to \mbox{\textit{super-ETI}} as ETI $10^3+$ years more advanced than us.  Exploration undertaken by super-ETI would not only utilize probes of vast intelligence, but would greatly benefit from life-like self-replication mechanisms.  Exploring the galaxy solely with resources mined from, and machinery built in, the homeworld solar system is a tremendous energy and economic drain.  Self-replication solves this problem by offloading the energy and economic expenditure to the progressing mission itself.  For the mere cost of an initial generation of probes, the entire galaxy can be explored.  The feasibility of self-replicating machinery is a difficult proposition for many people to accept, but life itself is proof-positive that matter and energy can be organized in such a fashion.  The notion that super-ETI could not artificially emulate life-like processes is completely absurd.  Life is admittedly very complicated, but it is not \textbf{\textit{that}} complicated!

SRPs dramatically decrease the exploration time.  While colonization efforts might be slow (a point we reconsider in section \ref{label_section_AFullyComputerizedCivilization}), intentional self-replicating exploration ahead of the colonization wave has been shown to be extremely fast: \mbox{$4$$\times$$10^6$}--\mbox{$3$$\times$$ 10^8$} years is sufficient to fill the galaxy \citep{Hart:1975,Tipler:1980,Jones:1981}.  Thus, they ought to be here already, and many times over at that, as shown in section \ref{label_section_EstimatingtheSRPPopulation}.

The debate over SRPs has gained a certain degree of infamy.  Two seminal papers have dominated the literature, one by Tipler, espousing SRPs \citep{Tipler:1980} which extends an argument by Hart \citep{Hart:1975}, and one opposing Tipler by Sagan and Newman \citep{SaganNewman:1983}.  On the basis of this debate, and perhaps due to a strong desire for ETI to exist in the face of paradoxical theories, recent models of galactic exploration have explicitly excluded SRPs, citing these (and other) papers as their reasoning.  In other words, an isolated thirty-year-old argument has redirected the entire field and shuttered possible avenues of research.  The field of Fermi Paradox research would benefit if explanations other than the exclusion of SRPs were considered.

One point we take issue with is an inherent and frequently unconscious \textit{biological bias} that pervades consideration of computerized intelligence, including SRPs.  We expand on this idea in sections \ref{label_section_BiologicalBias} and \ref{label_section_AFullyComputerizedCivilization}.

In section \ref{label_section_EstimatingtheSRPPopulation} we estimate the number of SRPs in our solar system.  Then, in sections \ref{label_section_SelfReplicatingProbeVoluntaryRefrainment} and \ref{label_section_SelfReplicatingProbeInefficacy} we consider two arguments against SRPs, one by Sagan and Newman which we call  the \textit{voluntary refrainment} argument and one by Chyba and Hand which we call the \textit{inefficacy} argument.  In both cases, we believe there is opportunity to revisit their conclusions and that the total abandonment of SRPs from recent models of galactic exploration is premature.

\subsection{Estimating the SRP Population}
\label{label_section_EstimatingtheSRPPopulation}

To investigate how many SRPs have reached our solar system, we can use an equation which mirrors the method employed by the Drake equation \citep{Brin:1982, Walters:1980}, \ie by combining fractional parameters and a starting value.  Thus, on the assumption that each SRP mission targets one SRP per star, the number of SRPs that reach each star in the galaxy is \mbox{$N_r=N_s\cdot f_r\cdot n_r\cdot G_r$} where $N_s$ is the total number of stellar societies to ever arise (including colonies), $f_r$ is the fraction of societies that dispatch SRPs, $n_r$ is the number of missions they initiate, and $G_r$ is the fraction of the galaxy that each mission reaches.

There are virtually no estimates of $N_s$ in the literature; most models either derive the \textit{current} ETI population, \eg \citep{CottaMorales:2010} and any application of the Drake equation, or disregard the effects of interstellar colonization, \eg \citep{Forgan:2009, Bjoerk:2007, CottaMorales:2010}.  A broad overview of the literature suggests a consensus for the \textit{total} number of \textit{civilizations} (species) to be \mbox{$3$$\times$$10^2$}--\mbox{$10^{10}$}, but common estimates practically never fall below \mbox{$10^4$}.  Likewise, we can rightly add at least an order of magnitude to account for colonization.  We will assume \mbox{$10^5 \le N_s \le 10^{10}$}.

$f_r$ should be quite high; once SRP technology is available, its advantages are too significant to ignore.  In section \ref{label_section_SelfReplicatingProbeVoluntaryRefrainment}, we consider a counter-argument, but find it unsatisfactory.  $f_r$ is primarily impeded by the mean stellar society lifetime, $L_s$ ($\sim$$L$ from the Drake equation but $L_s$ applies to stellar societies, not species) (see sections \ref{label_section_PercolationTheorywithColonyDeath} and \ref{label_section_RefiningtheDrakeEquation}).  However, SRPs are practically immune to societal death.  By definition, SRPs can continue their mission even if their society dies.  If humanity's technological progress is any indication, SRPs should be feasible within $\sim$200-500 years of technological ascendancy (assuming we started \mbox{$\sim$100} years ago), thus invalidating all but the briefest estimates for $L_s$.  Furthermore, secondary colonies should produce SRPs \textbf{\textit{much}} more quickly than the homeworld since they are already technologically advanced.  We will assume \mbox{$f_r  \ge 0.1$}.

The purpose of $n_r$ is to emphasize that over centuries or millennia, societies may easily undertake otherwise redundant projects.  We will assume \mbox{$1 \le n_r \le 10$}.

$G_r$ is arguably very high.  In section \ref{label_section_SelfReplicatingProbeInefficacy}, we consider a counter-argument, but find it unsatisfactory.  As a concession, we will assume \mbox{$G_r \ge .01$}, but there is little justification for such deficiency.

When we populate the equation with the lower and upper estimates, it yields \mbox{$N_r=10^2$}--\mbox{$10^{11}$} SRPs in our solar system at this very moment.  The absurdity of this result underlies the tremendous burden of the Fermi Paradox and demonstrates why many ETI-hopefuls have shied away from even the remotest consideration of SRPs.

\subsection{Self-Replicating Probe Voluntary Refrainment}
\label{label_section_SelfReplicatingProbeVoluntaryRefrainment}

One popular argument against SRPs is presented by Sagan and Newman \citep{SaganNewman:1983}.  They argue that any presumably wise and cautious civilization would never develop SRPs because such machines would pose an existential risk to the original civilization.  The concern is that the probes may undergo a mutation which permits and motivates them to either wipe out the homeworld or overcome any reasonable limit on their reproduction rate, in effect becoming a technological cancer that converts every last ounce of matter in the galaxy into SRPs.

Ironically, one of the best counter-arguments to the Sagan/Newman theory is described in Tipler's original paper to which they were responding.  Tipler conceded the danger of mutation and unintended behavior or reproduction.  He pointed out that unlike biology, which is subject only to the limitations imposed by unintentional and unconscious evolution, machines designed with foresight and intelligent engineering may incorporate fundamental restrictions on erroneous reproduction.  Such restrictions could be so deeply ingrained as to render any mutated individual a still-birth (a simple checksum might suffice).  We know that engineering can achieve remarkably high data integrity rates.  Consider that each individual bit in a computer's dynamic memory (amongst billions) is not merely in stasis, but rather constantly leaks (it is a capacitor).  The computer must recharge every bit thousands of times per second (while also fixing radiation damage), with virtually no room for undetectable or uncorrectable errors.  The number of times this process occurs in all the world's computers in a single second is beyond comprehension, yet the worldwide undetected-uncorrected-RAM-bit-error rate is negligible (note that transient errors do occur, but the overall engineered process generally rectifies them).  More complex processes, such as file duplication or transmission, exhibit equally impressive consistency.  The sheer number of bits being duplicated and transmitted around the world in any given second is testament to the ability of purposeful engineering to achieve unfathomable data integrity rates.  To be clear, the point is not that super-ETI would use anything remotely resembling 20th century technology in SRPs; the point is that engineering as a general method for designing and constructing complex devices can achieve extremely high data integrity rates.

In addition to using our own technological experiences as a reference, we can consider a biological analogy.  Sagan and Newman are saying that detecting and correcting effectively cancerous events is not a safe bet on scales approximating galactic exploration.  The number of replication events in a galactic exploration mission, $R$, closely approximates the number of stars in the galaxy, $N$, so \mbox{$R \simeq N \simeq 100 \text{ billion}$} (within an order of magnitude).  The human body undergoes $\sim$10,000 trillion cell divisions over the course of its lifetime \citep{Turner:1978} or \mbox{$\sim$100,000$\times R$}.  Biology employs numerous  tactics to stem cancerous events, including detection of deleterious cells, repair of damaged DNA, and ultimately apoptosis if necessary.  Many humans go their entire lives without experiencing a single malignancy.  Of the cancers that do occur, many are benign, suggesting that not every cancerous event must necessarily yield unmitigated reproduction.  Lastly, and most crucially, biology is the product of an unconscious evolutionary process with no foresight or intelligent planning.  At the very least, we should expect intentional engineering to achieve biology's level of success, and more likely, to vastly exceed it.  In summary, $R$ is simply not all \textbf{\textit{that}} large a number compared to the biological equivalent and considering biology's success rate against cancer, and especially when considering intentional engineering.  Even if we discard engineering, we should still expect a galaxy-scale SRP cancer risk $\sim$1/100,000th that of an individual human.  Taking engineering into account could easily improve the situation by many orders of magnitude.

\subsubsection{Biological Bias}
\label{label_section_BiologicalBias}

Another counter-argument to Sagan and Newman's concern is to question the presumption that computerized intelligence must inherently be susceptible to such deficiencies in the first place.  We refer to this as \textit{biological bias}, the belief that computers cannot possess the intellectual power, mental generality, or cognitive richness of humans.  As with many forms of bias, it often goes unnoticed by its own purveyors -- they may happily grant the concept of machine intelligence for the sake of argument without consciously realizing that their subsequent criticisms imply machines which nevertheless lack human-level consciousness and self-actualization.  Fears of cancerous mutations that hyperbolically lead to the complete matter-assimilation of the cosmos (a scenario literally described by Sagan and Newman) are in accordance with biological bias and underly one of the primary arguments against SRPs.  While these are reasonable criticisms of contemporary human technology, there is no justification short of prejudice to extend them to super-ETI.  If we are to truly consider the nature of an advanced SRP, then we must grant that probe a versatile and brilliant intellect comparable to -- more likely vastly surpassing -- that of a human; it should possess deep powers of introspection and investigation and should easily recognize when one of its offspring is severely and harmfully abnormal.

We believe that much of the disagreement over this matter stems from a difference of philosophical stance.  Consider three scenarios under which an SRP cancer could occur:

\begin{enumerate}
\item A malevolent ETI deliberately releases a cancerous SRP.

\item An ETI slightly more advanced than us recklessly incorporates mechanized self-replication into a space-probe otherwise comparable to our current probes.

\item A super-ETI creates a brilliant and intellectually embodied SRP for the purpose of benevolent galactic exploration.
\end{enumerate}

We can dismiss the first scenario since as it is not in the minds of either side of this particular debate.  Of the remaining two scenarios, the Tipler camp envisions the third while the Sagan/Newman camp envisions the second.  The disparity between those two scenarios underlies the discord over this issue.  We recommend considering human cognition to be a lower-bound on the intellectual capacity not only of super-ETI, but of their computerized machinery as well.  As a litmus test for detecting potentially unconscious biological bias we propose the following:

\begin{quote}
\textit{If we cannot imagine the Sagan/Newman cancer scenario applying to a group of brilliant and wise humans exploring the galaxy, then we should dismiss it out of hand for SRPs as well.}
\end{quote}

Clearly, this test can be generalized to any argument against computerized intelligence.

\subsubsection{A Fully Computerized Civilization}
\label{label_section_AFullyComputerizedCivilization}

Another point is the presumed distinction between the SRP exploration phase and the subsequent populace-colonization phase.  One notable difference is the assumption that SRP exploration would proceed much more rapidly than colonization.   Given recent theories that technological species may evolve into a computerized intelligence, this distinction evaporates \citep{Shostak:2010,Wiley:2011}.  Exploration and colonization missions become one and the same; they even ride the same starships.  Here is a possible scenario: a ship travels at maximum velocity regardless of an exploration or colonization mission.  Upon arrival, while nascent colonization begins to populate the new solar system, construction ($\sim$ self-replication) of the next generation of ships commences immediately and simultaneously along-side the nascent colony.  The new ships are launched just as quickly as in the original SRP-exploration scenario.  In this way, a computerized civilization might simultaneously explore and colonize the galaxy at an astonishing rate.

If this scenario offers any insight whatsoever into plausible methods of galactic colonization, it has dire implications for the Fermi Paradox.  Not only would SRPs fill the galaxy very rapidly (an event which we might miss if the probes undertake minimal planetary reorganization), but now full galactic colonization might proceed at a similar rate due to the computerization of a civilization's primary population.  We will return to the notion of computerized intelligence when we consider recent models of galactic exploration by Bj$\o$rk, and Cotta and Morrales.

\subsection{Self-Replicating Probe Inefficacy}
\label{label_section_SelfReplicatingProbeInefficacy}

Chyba and Hand \citep{ChybaHand:2005} offer a completely different argument against SRPs: that the efficacy of SRPs (their expansion rate) would be intrinsically truncated by the occurrence of mutated predators.  The postulated mutants would discard their original imperative of exploration to prey upon the remaining nonmutated probes.  Such behavior would greatly impede the efficiency of the original mission and would undermine claims of rapid SRP exploration.  In an extended version of their argument a complex ecology (a food web) of SRPs might arise and disband from the original mission to deal with the newly evolved business of chasing and eating one another in some confined corner of the galaxy.  Chyba and Hand's reasoning is fundamentally different from that of Sagan and Newman.  The latter argue that no civilization would bother to create SRPs in the first place.  Chyba and Hand argue that even if attempted, the presumed expansion rate would be greatly impeded by predation.

Earthly examples inspire confidence in their theory.  Predators can have a drastic impact on a prey's population.  However, one unspoken and numbingly obvious reason that this is possible is that the predators can actually \textbf{\textit{catch}} the prey.  Furthermore, the predation rate must approach the prey reproduction rate if the prey's population growth is to be noticeably impeded.  Natural ecosystems display these patterns because the prey exhibit a few notable properties:

\begin{enumerate}
\item They don't spend their entire lives dispersing radially from a locus of universal origin.

\item They don't always travel at maximum speed.

\item They periodically rest, sleep, and eat.
\end{enumerate}

These properties permit predators to physically catch up with the prey.  Yet they are unlikely to represent a galactic exploration scenario.  First of all, we should assume that interactions are only possible in solar systems.  That is to say, one probe cannot easily chase down another in the lonely void between stars, much less at a tenth the speed of light!  The initial probes are dispatched in a radial pattern away from the homeworld at the maximum interstellar speed.  In each solar system they self-replicate as quickly as possible and dispatch offspring into an approximate hemisphere oriented away from the homeworld.  This period of regeneration offers predators a chance to descend upon them.  However, it does not indicate an increase in the relative predator/prey speed since predators must, of course, undertake the same respite to self-replicate.  An alternate scenario in which the predators don't self-replicate as often might permit them a briefer regeneration period, thus enabling them to overtake the prey...but only in one direction -- the distribution of the nonmutated probes renders this strategy incapable of culling their numbers in all directions at once.

These behaviors make it exceedingly difficult for the predators to overtake and wipe out the prey.  The dispersal pattern is not terribly unlike that of a bacterial culture growing in a petri-dish.  In fact, to the extent that the galaxy is quasi-planar, the analogy is that much more apt.  If we treat the initial mission as a disk expanding outward from the homeworld, and if we subsequently treat the point mutation of a predator as the origin of a new expansion disk (perhaps only expanding back into the original disk, but not outward), then it is clear that the predator disk is always smaller than the prey disk -- the predators can never hope to overtake and wipe out the nonmutated probes in an infinite space.  However, petri-dishes and galaxies are not infinite and this definitely impacts the model.  For example, if we assume that the predators are perfectly lethal \textbf{\textit{and}} that the prey are perfectly defenseless (and why \textbf{\textit{should}} we assume this anyway?!) then the predators will eventually wipe out the mission-mindful probes.

\begin{figure}[t]
\centering
\includegraphics[width=\linewidth]{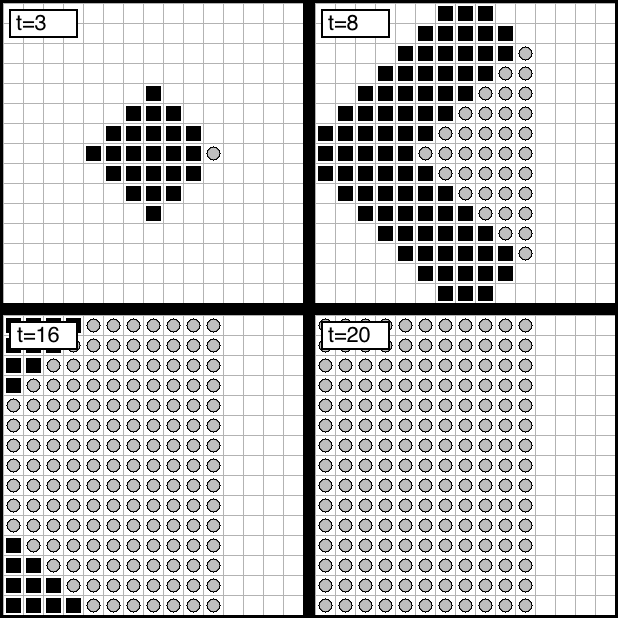}
\caption{\textbf{Predator Shadows (Dark squares: healthy probes, Light circles: predators):} While predators cannot overtake distant prey until the boundary is reached (\eg left side of the board), they can nevertheless produce a shadow of unexplored territory (\eg right side of the board).}
\label{figPredatorShadows}
\end{figure}

\begin{figure}[t]
\centering
\includegraphics[width=\linewidth]{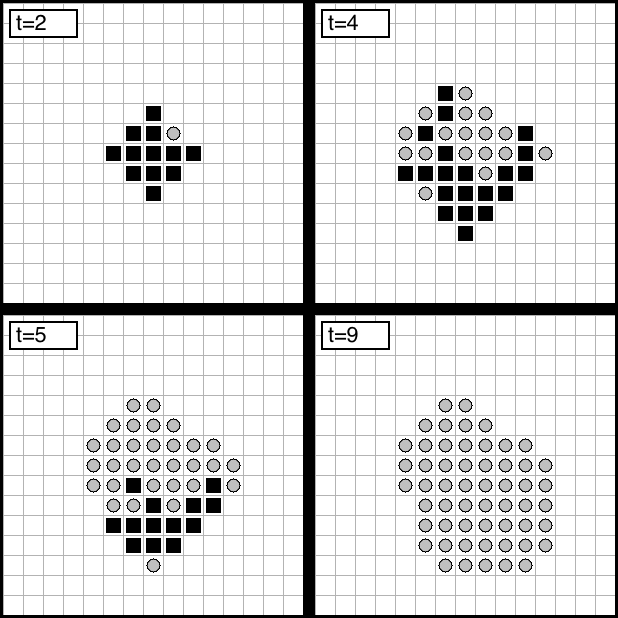}
\caption{\textbf{Predator Confinement (Dark squares: healthy probes, Light circles: predators):} If a sufficient number of mutated predators emerge, they can finitely confine the civilization's expansion.}
\label{figPredatorBounded}
\end{figure}

Two questions then arise: can predators prevent the nonmutated probes from reaching the far boundary, and if not, can predators prevent any other region of the galaxy from being reached, thus leaving voids in potential contact and detection?  The answer to the former question is no.  Clearly, if predators arise east of the homeworld (relatively speaking), they can never catch the wavefront of probes expanding westward before those probes reach the west edge of the galaxy.  However, some areas of the galaxy may never be visited, namely the region beyond the frontier of the predators.  Thus, we can speak of a \textit{predator shadow}, a region of the galaxy which is not reached by that particular ETI civilization (see Fig. \ref{figPredatorShadows}).  Furthermore, if there are a sufficient number of independent predatory mutation sites surrounding the homeworld, they may even bound the nonmutated probes' expansion.  In this way, it is possible to enable Chyba and Hand's model to finitely confine a civilization's reach via SRPs (see Fig. \ref{figPredatorBounded}).  This would seem to be a concession to Chyba and Hand's point, but it suffers two weaknesses.

First, we must consider a scenario involving numerous independent ETI distributed throughout the galaxy and time.  Even if each civilization is bounded by an outer shell of predators, the union (overlap) of their reach may still fill the galaxy.  Furthermore, if there are not enough predatory mutations to fully confine each civilization, but merely enough to shadow them, then far fewer independent ETI homeworlds are necessary to fill the galaxy; a handful should suffice.

The second weakness of this argument was previously stated in section \ref{label_section_SelfReplicatingProbeVoluntaryRefrainment}.  We must take some realistic account of the likelihood of the mutations that Chyba and Hand are positing, especially at sufficient rates to fully bound expansion as opposed to merely shadow it.  As described, it is possible to achieve data replication rates which render $R$, the number of replication events for galactic exploration, quite safe.  Thus, while predatory mutations may not be fundamentally impossible, let us not overestimate their risk when considering the efficacy of SRPs.

\subsection{Validity of Models that Exclude Self-Replicating Probes and Assume Slow Colonization Rates}
\label{label_section_ValidityofModelsthatExcludeSelfReplicatingProbesandAssumeSlowColonizationRates}

Two impressively detailed models of galactic exploration are described by Bj$\o$rk \citep{Bjoerk:2007}, and Cotta and Morales \citep{CottaMorales:2010}.  Both models explicitly exclude SRPs and cite Sagan and Newman, and Chyba and Hand, as their primary reasons for doing so.  The weaknesses of the Sagan/Newman and Chyba/Hand arguments therefore directly compromise Bj$\o$rk and Cotta/Morales' results; their models may deliberately incorporate handicaps based on unreasonable assumptions.  A third model, by Forgan \citep{Forgan:2009}, is at best vague about the effects of SRPs in that it makes practically no mention of them.  In fact, Forgan's current model explicitly excludes any form of interstellar migration.

Furthermore, given the theory described in section \ref{label_section_AFullyComputerizedCivilization} that colonization by a computerized society might proceed just as rapidly as exploration, the fact that both Bj$\o$rk and Cotta/Morales also exclude colonization renders their models potentially less realistic on that point as well.  Namely, realistic models of galactic exploration and colonization should consider not only the likely scenario of SRPs, but also the possibility that colonization itself may proceed at a phenomenal rate.

We conclude that Bj$\o$rk and Cotta/Morales's models are unlikely to inform us about realistic circumstances of galactic exploration and colonization.  While Forgan's model is also incomplete with respect to both SRPs and colonization, he specifically emphasizes the model's flexibility and potential for extension in the future.  We look forward to it.

In this section we summarized our thoughts on the Fermi Paradox with reference to the specific subtopic of SRPs.  In the next section we switch gears to consider the Fermi Paradox from the point of view of percolation theory.

\section{Percolation Theory}
\label{label_section_PercolationTheory}

Landis explores the Fermi Paradox by considering a model based on percolation theory \citep{Landis:1998}.  Each new colony is created in one of two states: \textit{colonizing} or \textit{noncolonizing}, the reason being that while some cultures opt toward exploration and expansion, others develop nonexpansionist values and display minimal interstellar reach.  In the model, only colonizing colonies propagate offspring to neighboring sites.  Each new colony is assigned one of the two states with some predetermined probability, $P$.  Landis uses \mbox{$P=1/3$}.  The only other factor which impacts the behavior of a percolation system is the connectivity or \textit{degree} of the graph connecting the sites, $N$.  Landis argues in favor of \mbox{$N$ $\simeq$} 5 for natural interstellar neighborhoods, although in the one example he shows he uses \mbox{$N=6$} because the simulation resides in a cubical packing.

Landis explains that given the two parameters, $P$ and $N$, there exists some critical value, $P_c$, such that for \mbox{$P < P_c$}, colonization inevitably peters out with an exterior shell comprised solely of noncolonizing colonies which entrap any colonizing colonies.  Thus, civilizations are stochastically bounded to some finite region of the galaxy.  If this property holds for virtually all civilizations, and if the number of independent civilizations is low enough that their union (overlap) does not fill the galaxy, then this theory speaks to the Fermi Paradox.  Even for \mbox{$P > P_c$}, in which the percolation model permits civilizations to grow indefinitely, thus spanning the galaxy, there still occur inverted shells whose surfaces comprise solely of noncolonizing colonies such that their interior regions are never visited.  Thus, even for large values of $P$ it is still possible, albeit less likely, for Earth to reside in an inaccessible region.  Obviously, the larger $P$ is, the fewer independent ETI are required such that their union is likely to be complete, perhaps as few as two.

Percolation theory offers a tantalizing explanation for the Fermi Paradox.  One of its strongest assets is that it minimizes our reliance on speculative or whimsical musings about the sociological motives and behavior of alien species.  Rather, it suggests that there may be fundamental graph-properties of expansion which describe galactic exploration in purely mathematical terms.  It is still somewhat speculative however, in that we must choose $P$, and to a lesser extent, $N$.

\begin{figure}[t]
\centering
\includegraphics[width=\linewidth]{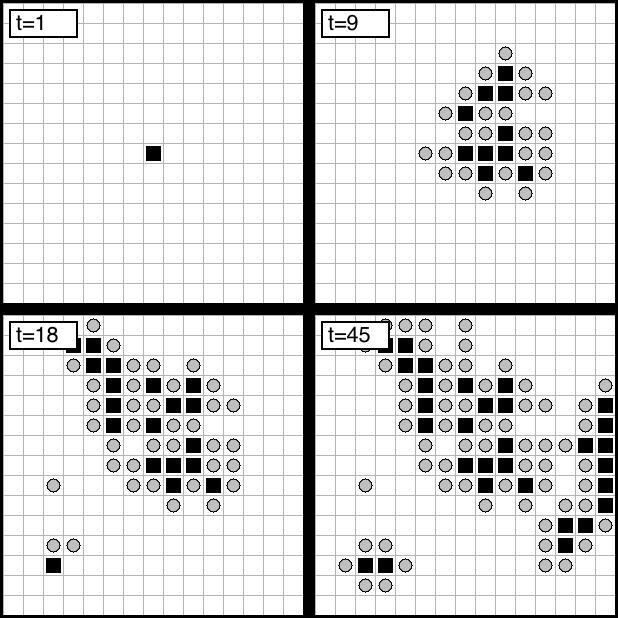}
\caption{\textbf{Landis' Percolation Model (Dark squares: colonizing, Light circles: noncolonizing):} For sufficiently small values of $P$, colonization eventually halts with all colonizing colonies trapped within a shell of noncolonizing colonies.  In the figure, \mbox{$N$=$6$} (the figure depicts one layer of a cubical simulation) and \mbox{$P$=$1/3$}.  We observe that every colonizing colony becomes trapped and thus no further expansion is possible.}
\label{figPercolationOriginal}
\end{figure}

A pure percolation model does indeed split the galaxy into distinct regions, some of which are never visited (see Fig. \ref{figPercolationOriginal}).  However the model is extremely simple and may not reflect realistic scenarios.  Landis concedes this point but argues that while extensions to the basic model could certainly be envisaged, they would not be likely to alter the fundamental property that certain regions remain unvisited.  This assumption is incorrect however.  There are perfectly reasonable extensions to Landis' model which completely alter its behavior.  In sections \ref{label_section_PercolationTheorywithColonyDeath} and \ref{label_section_PercolationTheorywithColonyStateMutation} we introduce two such extensions and show how they affect the model.

\subsection{Percolation Theory with Colony Death}
\label{label_section_PercolationTheorywithColonyDeath}

\begin{figure}[t]
\centering
\includegraphics[width=\linewidth]{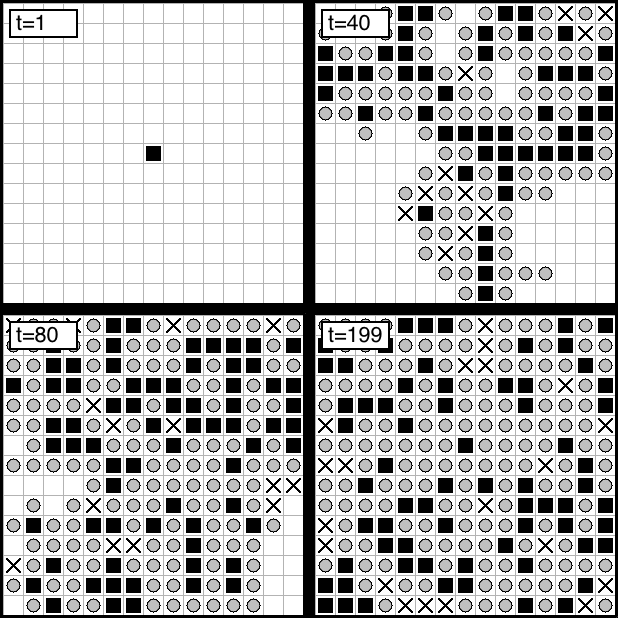}
\caption{\textbf{Percolation Model with Periodic Colony Death (Dark squares: colonizing, Light circles: noncolonizing, Xs: vacant but visited):} For the exact same value of $P$ which yielded a halting expansion in the original model (see Fig. \ref{figPercolationOriginal}), we can achieve saturation by introducing $K$, a colony's lifetime.  Death on the entrapping shell permits trapped colonizing colonies to leak through and continue the colonization effort.  (\mbox{$N$=$6$}, \mbox{$P$=$1/3$}, and \mbox{$K$=$20$}).}
\label{figPercolationWithDeath}
\end{figure}

From the very beginning, the Drake equation has included the parameter $L$, the lifetime of a technological civilization.  The expectation that advanced civilizations may die is a widely accepted idea in the SETI community (popularly credited to our own nuclear standoff) and therefore is a perfectly reasonable addition to any model of galactic colonization.  Introducing an analogous \textbf{\textit{colony}} lifetime, $K$, to Landis' model implies that colonies periodically die.  Notice that $K$ differs from $L$ in that $L$ refers to entire civilizations, not individual colonies.  When a colony dies on the frontier, it opens a valve in the otherwise impenetrable shell through which trapped colonizing colonies can leak out.  Since the point of the percolation model is that certain locations are \textbf{\textit{never}} visited, and since this leakage occurs in the outward direction, this modified model is now fundamentally different -- it permits the expansion process to potentially reach every star in the galaxy (see Fig. \ref{figPercolationWithDeath}).  It is not guaranteed however.  Now that colonies can die, the entire civilization might \textit{evaporate} before it reaches every locus.  Much as $P_c$ defined a threshold which differentiated between bounded and indefinite expansion, we can now speak of a threshold $K_c$, such that for \mbox{$K < K_c$}, the civilization evaporates before filling an infinite space, and for \mbox{$K > K_c$}, the civilization may span the space (and even fill the inverted shells previously described).  Given that $K_c$ varies with $P$ (larger values of $P$ permit lower values of $K$ without evaporation) we can state it more precisely as $K_{Pc}$, the threshold for a given $P$.

The notion of $K$ speaks to a broader philosophical issue when considering the nature of ETI, which is that perhaps our conceptualization of $L$ is not the correct property of interest in regard to the Drake equation's intended purpose.  If the Drake equation is used to count the number of detectable \textit{species}, then so be it, L is appropriate, but if the intended use is to count the number of culturally distinct societies or to estimate the odds of SETI detection, then perhaps we should count \textit{stellar societies}, not species (an idea previously proposed by both Brin and Walters et. al. \citep{Brin:1982, Walters:1980}).  If every colonized solar system is regarded as a separate ETI event that may make itself detectable via either radio or visitation, we can effectively replace $f_c$ (the fraction of intelligent species that communicate) with a new parameter $n_s$, the number of stellar societies per species, including the homeworld.  We expand on this idea in section \ref{label_section_RefiningtheDrakeEquation} when we offer a revised version of the Drake equation.

\begin{figure}[t]
\centering
\includegraphics[width=\linewidth]{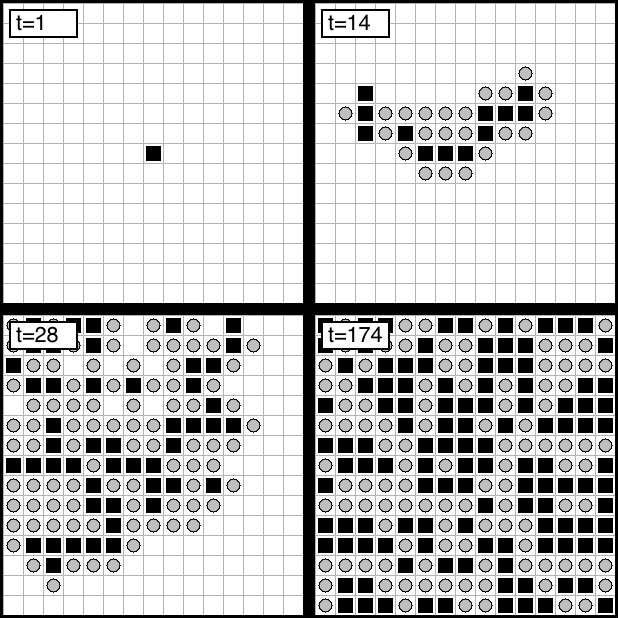}
\caption{\textbf{Percolation Model with Periodic Colony Mutation (Dark squares: colonizing, Light circles: noncolonizing):} For the exact same value of $P$ which yielded a halting expansion in the original model (see Fig. \ref{figPercolationOriginal}), we can achieve saturation by introducing $M$, the probability that a colony will mutate its colonizing state in a given timestep.  Mutation on the entrapping shell permits noncolonizing colonies to become colonizing and restart the colonization effort from their location.  (\mbox{$N$=$6$}, \mbox{$P$=$1/3$}, and \mbox{$M$=$.01$}).}
\label{figPercolationWithMutation}
\end{figure}

\subsection{Percolation Theory with Colony State Mutation}
\label{label_section_PercolationTheorywithColonyStateMutation}

In addition to introducing the concept of colony death, we have added $M$, a mutation rate in the colonizing state of a colony.  M permits any colony to switch its state at any time with some low likelihood.  The justification for such a parameter seems quite reasonable: it is absurd to assume that colonies will remain culturally fixated on either of the two motives (colonizing or noncolonizing) forever.  Societies evolve, beliefs and values vacillate, cultures meander.  $M$ permits currently noncolonizing colonies on the frontier to periodically mutate and consequently promote a new wave of exploration outward from their location, once again undermining the conclusion of the original percolation model (see Fig. \ref{figPercolationWithMutation}).  Furthermore, note that by combining both of the new parameters, $K$ and $M$, our simulations confirm that we can achieve saturation more quickly than with either parameter alone for a given $P$.

The point isn't merely that $K$ and $M$ permit civilizations to reach a little further or last a little longer.  The point is that the fundamental property of the original model is overturned.  Given enough time, \mbox{$K > K_{Pc}$} or any nonzero value of $M$ permit every locus in the simulation to be reached -- 100\% saturation is guaranteed.  The implication of our modified model for the Fermi Paradox is obvious: it greatly eases the task of exploring the galaxy without being bounded by a shell of noncolonization.

Ultimately, simplistic models such as Landis' and our variant cannot hope to perfectly represent a topic as complex as galactic colonization.  We do not use such models to draw final conclusions about the uncompromising truth of reality.  Rather, we ask which models are most likely to indicate natural circumstances with the greatest verisimilitude.  We believe that where the phenomenon under investigation is the supposition that shells of noncolonization form permanently impenetrable barriers, our modified model more closely resembles reality than Landis' original model.

In this section we summarized our thoughts on Landis' adaptation of percolation theory to galactic colonization and its implications on the Fermi Paradox.  In the next section we introduce ITB theory and consider its implications for theories of interstellar societal collapse.

\section{Interstellar Societal Collapse}
\label{label_section_InterstellarSocietalCollapse}

Some researchers have considered the Fermi Paradox from the perspective of interstellar societal collapse.  We introduce a new theory of \textit{interstellar transportation bandwidth} (ITB) which may impact such theories.  We begin with a survey of two such papers and present some reasonable challenges in the absence of the new theory.  We then introduce the ITB theory and consider its additional implications for theories of interstellar societal collapse.  Later, we show how ITB theory can be used to refine the Drake equation so as to produce more precise estimates of the intragalactic ETI population, and finally show how to calculate the ITB.

\subsection{Light Cage}
\label{label_section_LightCage}

McInnes shows that an expanding sphere of colonization cannot expand fast enough to avoid complete societal collapse due to exponential population growth and increasing population density \citep{McInnes:2002}.  Furthermore, after such a crash, the society may be so impoverished of resources that it is forever blocked from making a second attempt.  If this theory is universally true of all civilizations, then it offers significant insight into the Fermi Paradox.  McInnes extends the intuitive notion of a constant expansion rate with a steadily increasing population density to a linear expansion rate which maintains constant population density.  However, the expansion velocity must have some maximum speed limit.  To generate an upper-bound for this theory, McInnes considers a maximum expansion velocity of 1.0c.  Obviously, in a real scenario, the expansion rate would be considerably slower, but that only exacerbates McInnes' fundamental point which is that once the maximum expansion velocity is reached, population density once again begins to rise.  Eventually, total societal collapse is inevitable.  The only way to avoid catastrophe is to shift from exponential growth to logistic growth.

Given a specified growth rate and an initial sphere of some specified size, one can calculate the size of the expanded sphere when the increasing expansion rate reaches the 1.0c limit.  McInnes calls this sphere the \textit{light cage}.  Assuming 1\% annual growth and a starting  sphere the size of Earth, he determines that the light cage is at a 300ly (light-years) radius and that the time to reach the light cage is 8000 years.  Out of curiosity, we ran the numbers for a more realistic maximum expansion velocity.  If we assume a maximum flight speed of 0.1c, a common estimate for fusion-powered starships such as Daedalus \citep{Daedalus:1978, Matloff:2005}, which yields voyages to nearby stars on the order of decades, and if we assume a regeneration period of a few decades -- and therefore comparable to the duration of the voyages themselves -- then we estimate a more realistic maximum expansion velocity to be \mbox{$\sim$.05c}.  When applied to McInnes' equations, this velocity reveals a far more realistic cage of a mere 15ly, and an associated saturation time of 7000 years.  Interestingly, while a considerably slower expansion rate drastically reduces the cage, it only slightly decreases the time until the cage is hit.  This follows naturally from exponential growth.  To make matters worse, .05c may be entirely too fast if the stasis period between voyages is longer than the few decades of the previous estimate.  If we briefly consider a maximum expansion velocity of .01c, we derive a cage of only 3ly!

Considering that there are only 52 stars within 15ly of Earth (and \textbf{\textit{none}} within 3ly!), and that only some fairly small subset will support habitation, this does not provide us with very many systems to colonize before we implode in a population density catastrophe.  More crucially, 15ly probably lies within the single-voyage horizon, which implies that early waves of colonization may fill the cage all at once as opposed to diffusing radially.  In other words, according to light cage theory, interstellar travel hardly provides any population relief at all, right from the beginning.

McInnes is careful to admit that his model is continuous and symmetrical.  Admittedly, galactic colonization is loosely like an expanding sphere, but more precisely it is like traversal of a rooted tree (a graph), where the root is the homeworld, vertices are solar systems, and labeled edges connect stars whose distances are below a maximum traversal threshold with the labels indicating interstellar distances (and associated travel times).  In addition to the inherently discrete nature of this graph, the natural distribution of stars will impose notable asymmetries on the edge length and per-node degree (branching factor).  McInnes admits to the simplicity of his model, but points out that such details should not affect the model's outcome.  We agree, but in practicality, a tree of no more than 52 nodes (probably far fewer), all of which have direct edges to the homeworld, no longer resembles a continuous sphere in even the weakest sense.  A Monte Carlo simulation might help illuminate any interesting properties of the discrete model.

Assuming that the continuous model is apropos to the discussion, then we envisage two ways by which the model may not predict actual events.  The first is admitted by McInnes, that populations may succeed in converting to logistic growth and therefore avoid the prescribed collapse.  Given the calculation of a 15ly cage and the observation that it would saturate in a single emigration wave due to direct homeworld access, any hope of survival would depend on adopting logistic growth prior to interstellar colonization; any later would be too late to avoid disaster.  However, Earth appears to be following just such a course, leveling off its population growth in the 21st century, long before undertaking interstellar travel \citep{MisraBaum:2009}.  Not only does this bode well for humanity, it suggests that other ETI civilizations can do the same...but such a conclusion weakens McInnes' theory that the Fermi Paradox is resolved due to a finite expansion before societal collapse.  If ETI are once again enabled to continue steady outward expansion, then the original questions underlying the paradox resurface.

The second challenge we would raise in response to McInnes' model will be explained in section \ref{label_section_InterstellarTransportationBandwidth} when we introduce ITB theory.  Briefly, we are unconvinced that interstellar distances don't impose an insurmountable paucity of interaction events between solar systems such that contagions of societal collapse fail to adequately infect neighbors.

Our final thoughts on McInnes' model reflect on his post-analysis.  He proposes the example of a civilization which permits exponential growth at its frontier but enforces logistic growth wherever resource limits are reached, \ie deeper within the sphere.  He shows how such a civilization could achieve phenomenal rates of unbounded expansion, filling the galaxy in a few millions years.  While McInnes' concedes this scenario, he nevertheless questions its validity by citing Landis' percolation model, namely to doubt that the civilization would expand indefinitely.  Considering that in section \ref{label_section_PercolationTheory} we demonstrate some weaknesses of the percolation model, we likewise conclude that the proposed scenario stands relatively unchallenged.  Should it turn out to be reasonable, then the Fermi Paradox remains, well, paradoxical.

\subsection{Unsustainable Expansion}
\label{label_section_UnsustainableExpansion}

Haqq-Misra and Baum present a theory which closely resembles McInnes' light cage, namely that practically all civilizations which are ravenous in their expansion behavior might collapse, leaving only slowly expanding civilizations \citep{MisraBaum:2009}.  To prove that societies capable of maintaining a slow expansion rate are possible, Haqq-Misra and Baum cite the Kung San of the Kalahari Desert.  However, we feel that this example proves the opposing point: the Kung San did not colonize and inherit the Earth; Western civilization did, precisely because of its rapid expansion.  Haqq-Misra and Baum propose that such civilizations may ubiquitously burn out, leaving only the slower civilizations.  While this may be a reasonable supposition, it is not strongly supported by our single datapoint, humanity.  After all, in Earth's case, the rapidly expanding civilizations won, in effect.  We do not live in a Kung San world precisely because they did not colonize it, and likewise, Western civilization did not burn out before colonizing a substantial portion of the planet.  Not only do we observe that rapidly expanding civilizations successfully dominated the planet (ethical considerations aside), but furthermore it looks as if the global human population will in fact level off in the 21st century (a point Haqq-Misra and Baum themselves cite), calling into question the secondary presumption that perhaps such civilizations must inevitably implode after reaching a planet's carrying capacity.  Thus, humanity's own history looks like a strong point of evidence \textbf{\textit{against}} sustainability theory.

If anything, this discussion suggests a natural selection favoring expanding civilizations.  Much as Western civilization overtook Earth at the Kung San's expense, perhaps we should expect rapidly expanding galactic civilizations to be the victors in the struggle for the galaxy.  This may very well be an imperative for survival, again using the Kung San and other indigenous tribes as an unfortunate example.  Perhaps only one civilization can dominate the galaxy while the rest must, if not go extinct, nevertheless lose considerable self-autonomy and cultural influence.  Ecologically, only one species can generally occupy a given niche at any place and time while, anthropologically, human history exhibits similar patterns in the struggles between colocated societies.  No doubt, as a progressive and intellectual species (one would hope), we can work actively against such forces, but even the need for such efforts is a blatant concession that the \textbf{\textit{natural}} progression of events is counter to our urbane values of equality and diversity.  This view is directly opposed to the suggestion that the most dominant and assertively expansionist societies should be the very ones we expect to fail.

Nevertheless, Haqq-Misra and Baum propose a fair question: whether rapid expansion can be sustained long enough to colonize the galaxy.  Perhaps the Earthly analogy is inappropriate in that rapid expansion is feasible on planetary scales but not on interstellar scales.  However, we are unsure \textbf{\textit{why}} such a supposition should be true.  Haqq-Misra and Baum support this claim by citing Landis and McInnes, but in sections \ref{label_section_PercolationTheory} and \ref{label_section_LightCage} we demonstrated possible weaknesses of both explanations.

Our strongest counter-argument to sustainability theory is described in section \ref{label_section_InterstellarTransportationBandwidth}.  Namely, our proposal of a possible ITB calls into question the ability of overpopulation, resource exhaustion, and other forms of societal strife to infect sufficiently remote societies.  If our theory is valid, then the only way sustainability theory can work is if societal collapse occurs prior to interstellar colonization...and furthermore, only if it occurs virtually ubiquitously across independent ETI homeworlds.  If sustainability theory only applies to some ETI before they embark upon interstellar colonization, and if ITB is itself a valid theory, then sustainability is intrinsically insufficient to resolve the Fermi Paradox.  The remaining civilizations should have colonized the galaxy in no uncertain terms.

\subsection{Interstellar Transportation Bandwidth}
\label{label_section_InterstellarTransportationBandwidth}

We believe the greatest challenge to theories of interstellar societal collapse is that it might be impossible for interstellar societies to contaminate one another with their respective problems, namely population and resource pressures, religious, socioeconomic, or political disputes, or other social strife which can cause complete societal disintegration.  There are three ways in which societies can interact, potentially imposing problems on one another:

\begin{enumerate}
\item Transportation of physical objects (colonists or otherwise) to a neighbor
\item Transportation of physical objects from a neighbor
\item Transmission or communication of immaterial data or ideas to a neighbor.
\end{enumerate}

In the first case, societal pressure is caused by the arrival of new colonists, and to a lesser degree, tangible goods which require either storage space or support.  In the second case, societal pressure may be a form of strip-mining which leaves the colony impoverished.  In the third case,  suffering might be transmitted through some remarkably destructive meme.  However, even if societally vanquishing memes are possible, such failure does not represent a sustainability pressure like population density or resource scarcity.  As such, it is not relevant to this discussion.

Interstellar resource extraction makes very little sense.  An important parameter in any space-travel equation is mass.  The mass of the various resources people require vastly exceeds the mass of the people themselves.  Consequently, it is always more efficient to move the people to the resources than vs/va, and by extension, even if interstellar hegemony and associated forced strip-mining is possible, it will not involve transport back to the homeworld, but rather, an invasion to use the resources where they reside, \ie we are back to the first scenario.

The arrival of unwelcome colonists is admittedly difficult for stellar societies to contend with because presumably magnanimous cultures cannot simply turn interstellar immigrants away; they will surely die after such a long trip if support is not provided very quickly.  In analyzing this pressure, we consider the \textit{interstellar transportation bandwidth (ITB): the number of people capable of moving from one solar system to another per unit time}.

We propose that the ITB is sufficiently low to shield each solar system from the pressures of its neighbors.  If this theory is valid, individual worlds might still collapse, but each colonized solar system effectively represents a unique experiment in the struggle against population growth and resource exhaustion.  For example, consider that the disaster of Easter Island's collapse did not infect any other society, including those of common origin, other Polynesian islands \citep{Diamond:2005}.  Easter Island was simply too isolated to transmit its failure elsewhere.  In fact, the inhabitants stranded themselves by deforesting to such a degree that they could no longer build ocean-going canoes.  By another analogy, extremely virulent diseases like Ebola do not necessarily wipe out enormous populations because the sick die too quickly to infect large numbers of people.  Consequently, in addition to our proposal of a bounded ITB in times of prosperity, perhaps during periods of degradation societies entirely lose the ability to sustain vast interstellar emigration (their ITB drops to zero).  Perhaps they can no longer build any canoes, as it were.

Another approach is to try to visualize how McInnes' propagating population pressure would actually occur.  The suggestion that an ever-rising population near the center of the interstellar civilization imparts an ever-increasing pressure on the frontier societies requires the central societies to \textit{exponentially} increase their starship production rate.  Not only is such a scenario seemingly illogical, it also violates the notion of an ITB.  More plausible scenarios would admittedly leave troubled worlds crushed by their own catastrophes, but would likewise shield other societies from accepting their burden.  Given that Haqq-Misra and Baum cite McInnes as one possible cause of their sustainability pressures, this point is germane to their argument as well.

One interesting consequence of the ITB is that for each new solar system reached, the odds of long-term survival by the species (survival by \textit{any} society within the collective) actually go \textbf{\textit{up}}, in direct opposition to McInnes' prediction.  This has a certain intuitive logic to it: each independent society increases the likelihood that someone somewhere survives.  A proverb involving eggs and baskets comes to mind.  It seems unlikely that both of these theories can be correct if one predicts an upward trend in survivability and the other predicts the opposite.

\subsubsection{Refining the Drake Equation}
\label{label_section_RefiningtheDrakeEquation}

ITB theory, and the related theory that societies undergoing a collapse may explicitly lose the ability to construct starships (the \textit{canoe theory} if we may be so bold), speaks more generally to the validity of $f_c$ and $L$ in the Drake equation \citep{Brin:1982, Walters:1980} (respectively, the fraction of intelligent-life-bearing planets that produce communicating civilizations, and the mean civilization lifetime).  As noted in section \ref{label_section_PercolationTheorywithColonyDeath}, a similar idea has been proposed by both Brin and Walters et. al. \citep{Brin:1982, Walters:1980}.  Perhaps each stellar society should be regarded as a separate entity, thus replacing $f_c$ with a new parameter, $n_s$, indicating the number of communicating stellar societies per species.  We can refine the meaning of $L$ as $L_\Omega$, the mean lifetime of the collection of societies originating from a single homeworld (from technological emergence on the homeworld to the death of the last associated society).  We may now consider a new related term, $L_s$, the mean lifetime of a single stellar society.  These new parameters, $n_s$ and $L_s$ replace $f_c$ and $L$ to produce an improved version of the Drake equation: \mbox{$N=R^*\cdot f_p\cdot n_e\cdot f_l\cdot f_i\cdot n_s\cdot L_s$}.

Consider how $f_c$ and $n_s$ differ.  Given a galaxy with ten planets surviving the $f_i$ filter (the fraction of life-bearing planets which produce an intelligent species), if only one of those species produces detectable technology, then \mbox{$f_c=0.1$}, but if that civilization colonizes two solar systems, the corresponding \mbox{$n_s=0.2$}.  Furthermore, $n_s$ can exceed 1, \eg if that one civilization colonizes twenty solar systems, then \mbox{$n_s=2$}.  Not only does $f_c$ fail to account for slight increases in colonization, as in the first example, but it is totally incapable of representing the second situation.  Interestingly, while $n_s$ is likely to exceed $f_c$, $L_s$ is likely to be less than $L$ (each society must last no longer than the collective civilization).  Thus, it is difficult to determine whether this new equation should produce a higher or lower value than the original (we're guessing higher)...but it should be more precise.

The number of stars within a single species' collective, $\xi$, scales cubically with radius out to the galactic thickness (\mbox{$\sim$1000ly}), and quadratically thereafter, and assuming a constant radial expansion rate, $v_e$, scales accordingly with time (incidentally, for \mbox{$v_e=.05\text{c}$} and a \mbox{$\sim$1000ly} galactic thickness, expansion undergoes this transition \mbox{$\sim$10,000yr} after initiation).  The number of habitable solar systems, $\xi_h$, is simply some fraction, $f_h$, of $\xi$ (where $f_h$ is some contortion of $f_p$ and $n_e$ from the Drake equation, so as to apply to stars rather than planets).  We can state $\xi$ (and therefore $\xi_h$) at a given time after technological emergence by applying the stellar density, $\rho$, to this sphere or cylinder (\mbox{$\rho=.004 \text{ stars}/\text{ly}^3$} near the sun, higher elsewhere \citep{Gregersen:2009}).  Taking into account $L_s$ and $r_g$ ($r_g$ = the mean regeneration period before dispatching new voyages), and assuming normal distributions, we can now estimate the probability that a stellar society dies childless, $P_\omega$, as the ratio of the \textit{normal difference distribution} (NDD) of $L_s$ and $r_g$ that is positive (\eg if \mbox{$r_g=L_s$} then the NDD is centered at 0 and the positive ratio = 0.5, but if \mbox{$r_g \gg L_s$} or \mbox{$r_g \ll L_s$}, then the NDD resides almost entirely to the right or left of 0, and thus the positive ratio \mbox{$\simeq$1} or \mbox{$\simeq$0}.  In the former case, the homeworld never spawns any offspring; in the latter, all societies produce offspring).

Furthermore, we can calculate the probability of total collapse at any given time as the binomial probability that all existing colonies die before dispatching new voyages: \mbox{$P_\Omega(t)=$}\mbox{${P_\omega}^{\xi_h(t)}$}.  It should be possible to extend this line of reasoning to deduce the expected time until a total collapse occurs, that is, solve for $t_\Omega$ such that \mbox{$\int_{t}^{t_\Omega} P_\Omega(t) \text{ d}t=0.5$}.  This would, in effect, yield $L_\Omega$.  This is a complex formulation however since $P_\Omega$ varies with so many other parameters including $t$ and $r_g$; it therefore lies beyond the scope of this paper.

Notably, the further a civilization expands, the greater its chances of indefinite survival.  Perhaps we can speak of a threshold $L_{\Omega c}$ (or an analogous threshold $L_{sc}$), such that for \mbox{$L_\Omega < L_{\Omega c}$} (\mbox{$L_s < L_{sc}$}), civilizations tend to evaporate and for \mbox{$L_\Omega > L_{\Omega c}$} (\mbox{$L_s > L_{sc}$}), civilizations tend to expand forever, consequently filling the galaxy.  In fact, the proposed model looks hauntingly similar to the modified percolation model presented in section \ref{label_section_PercolationTheorywithColonyDeath} with \mbox{$L_s \sim K$} and \mbox{$L_{sc} \sim K_{Pc}$}, such that $L_\Omega$ closely resembles the concept of evaporation presented in the percolation model.

\subsubsection{Calculating the Interstellar Transportation Bandwidth}
\label{label_section_CalculatingtheInterstellarTransportationBandwidth}

In this section, we show one  approach to calculating the ITB.  We loosely demonstrate our method with an example, but owing to the scarcity of research on some of the relevant parameters, it is difficult to be specific, or in some cases to even find reliable references.

Given a specified means of interstellar travel we can consider an associated crew size (colonist population) per ship, \eg a Daedalus starship \citep{Daedalus:1978, Matloff:2005}, modified to accommodate a manned mission.  Since Daedalus was never intended for manned missions, no crew estimates are available.  However, other designs offering comparable payloads (\mbox{$\sim$450} tons), such as Orion, support \mbox{$\sim$200} colonists (Orion was primarily intended for interplanetary travel which is why we aren't using it as our example).  The ability to dispatch ships is bounded by many factors such as construction materials (inconsequential \citep{Freitas:1983a}) and fuel, \eg 20K tons of deuterium and 30K tons of helium-3 for Daedalus \citep{Daedalus:1978}.  Besides the required quantity of materials, another relevant factor is time.  Daedalus is primarily time-bounded by the obtaining of scarce $^3$He.  The original design was for a twenty-year Jovian-atmosphere harvesting mission, although the schedule was amenable to scaling via concurrency (at an obvious corresponding scaling in cost).  In fact, another important consideration which we do not expand on here is economic limitations \citep{Dyson:1968,Daedalus:1978}.

With an estimate of the ship's crew size and the various production limitations, we can calculate the rate at which ships (and colonists) may emigrate, $C_\lambda$.  Our example yields one ship and 200 colonists every twenty years.  We could also estimate the total number of missions (colonists) the home solar system can physically support, $C_\Delta$, but this value is so large as to disregard (\eg Jupiter has $10^{16}$ tons of $^3$He, although perhaps heavy metals crucial to the ship itself would dominate such an analysis).

Dividing $C_\lambda$ by the number of neighboring stellar societies, $N_s$ (\eg \mbox{$N_s=4$}) yields the rate at which colonists can emigrate to a single colony, \mbox{$C_{s\lambda}=C_\lambda / N_s$} \mbox{$=(200/20\text{yr})/4$} \mbox{$=50/20\text{yr}$} (fifty colonists per system every twenty years).  In this fashion, we have defined an upper-bound on the \textit{colonist flux} that may be delivered to neighboring stellar societies and by which one society can potentially inflict its problems upon others.  We are now squarely in the realm of the intended analytical purpose of the ITB: whether it is sufficiently high for societal collapse in one solar system to detrimentally affect a neighbor.  To properly answer this question we would have to hear the input of sociologists and anthropologists.  Note that the ITB does not rely heavily on the flight speed.  Such a concept is akin to the notion of \textit{latency}, not bandwidth, and as such is less relevant to this discussion.

We have intentionally left this discussion quite vague.  Consider the effect upon the ITB by the type of starship.  Crucial parameters, such as the solar system's material and fuel supply, the crew size, and the rate of ship and fuel production, exhibited by various interstellar travel proposals should yield a wide range of ITB estimates.  For example, Super-Orion could transport several thousand colonists and was intended for interstellar missions \citep{Dyson:1968}, although its production timeline and economic limitations might amortize its long-term ITB relative to smaller ships.  It also seems that the final result of our example is almost unbelievable low.  It may be reasonable to permit super-ETI faster or concurrent ship-production and fuel-harvesting rates.  Of course, such benefits would come at increased cost (which must factor into any proper ITB calculation, despite its absence from our cursory example).  Furthermore, extremely advanced civilizations may span a wider interstellar collective, thus thinning the delivery to any individual neighbor.  Alternatively, if travel is concentrated unequally upon one colony (thus increasing the ITB), then we can increase the possibility of infectious societal collapse, but only by decreasing the ITB to the other colonies, thus undermining a theory of \textit{ubiquitous or total} collapse.  We will dispense with any further speculation on this matter, but the example shown here suggests that the ITB may be a very real bound on the transmission of interstellar societal pressure.

As an additional point of obfuscation, we could consider suspended animation or the transition to an entirely computerized species \citep{Wiley:2011, Shostak:2010}, as discussed in section \ref{label_section_AFullyComputerizedCivilization}.  Such methods could certainly increase the crew size of a starship.  Whether such increases would be sufficient for infectious societal pressures is uncertain.  In addition, with respect to a computerized species, the population of computerized individuals supportable by a given solar system would likely be greater as well and therefore the proportional impact of strife-ridden immigrants would not necessarily be any worse.

Most importantly, it is unclear how the concept of starship-population in human terms translates to that of an alien species.  This may be one of the most difficult aspects of ITB theory to resolve, aside from the general question of how societal pressure translates to an alien species' psychology and sociology.

The theory of an ITB exacerbates the Fermi Paradox by protecting each stellar society from the suffering of the others, thus mitigating theories of total civilization collapse based on a single point of origin from where harmful effects spread outward like an infection.  Total collapse is still a viable theory, but the likelihood of it occurring now becomes a binomial probability of independent collapse events.  Thus, with a steadily increasing expansion footprint, a homeworld's reach may easily fill the galaxy.

In this section we introduced ITB theory and considered its implications for the Fermi Paradox based on theories of societal collapse.  The rest of the paper considers one additional paper on the Fermi Paradox, then offers our thoughts on the strongest theory yet presented in the literature which may resolve the paradox without excluding the possibility of intragalactic ETI.  Finally, we offer our thoughts on how to better design future SETI programs.

\section{Bonus Stimulation}
\label{label_section_BonusStimulation}

Bezsudnov and Snarskii \citep{BezsudnovSnarskii:2010} present a model in which civilizations expand outward for a time, then succumb to parameter $L$ of the Drake equation and shrink back to their homeworld to eventually wink out.  However, if two civilizations collide along their expanding frontiers, they receive a \textit{bonus stimulation} (BS) to their lifetime.  In this manner, civilizations which meet other civilizations last longer, presumably due to the influx of new scientifically, technologically, and culturally invigorating ideas.

One possible weakness of this model is the manner in which civilizations die, \ie the youngest colonies die first, and so on back to the origin.  This model seems almost perfectly opposed to resource-limited models such as McInnes' which suggest that societal collapse propagates outward from the homeworld.  If we alter the BS model to die from the inside out, then the frontier colonies get a much longer time to meet other civilizations and receive the model's longevity reward.  Furthermore, it is unclear how the reward should affect societies within the civilization which have already died.  For both reasons, the dynamics of such a model are completely different from the original model.

More seriously, this model seems to disregard both the speed and the splintered nature of communication within an interstellar collective.  Consider the following peculiar behavior: if two civilizations, A and B, meet along their adjoining frontiers immediately prior to the moment when at least one of them, say A, would have begun to shrink, then both civilizations gain an immediate boost to their longevity.  Thus, the impending shrink is averted and both civilizations continue to expand, including on the opposite side of each civilization, maximally distant from the point of contact.  Contrarily, if the meeting had not occurred, then civilization A would have begun to shrink back to the homeworld.  Thus, the model posits an instantaneous (or at least temporally unresolvable) transmission of the BS.  Not only are we unconvinced by the near-instantaneous transmission of this salvaging effect across vast reaches of the galaxy, but more critically, we question whether any sort of cultural message of reinvigoration could maintain coherence given potentially thin channels of communication and visitation as per the ITB, especially considering the tremendously divergent societies that would have to pass the baton.

We do not mean to over-interpret Bezsudnov and Snarskii's model.  Surely they did not intend for it to be inspected at this level of precision.  Nevertheless, we propose that each stellar society be regarded as culturally unique.  While such an assumption may not seem reasonable in the initial decades after colonization, it is more reasonable for offspring societies separated from their parents by many millennia, with only intermittent and ITB-bounded physical visitation, and EM communication involving multi-decadal to multi-centennial delays.  This idea has a significant effect on the overall conceptualization of  galactic exploration and colonization; perhaps interstellar colonization is, in effect, a speciation event.

\section{ETI May Still Exist in our Galaxy}
\label{label_section_ETIMayStillExistinourGalaxy}

While arguments against SRPs are unconvincing, and while the other theories discussed may impose unrealistic assumptions, intragalactic ETI may yet exist.  The most convincing theory to date which permits intragalactic ETI is by Freitas who explains that exploratory probes could very well have reached our solar system and that we have overestimated the ease of detecting them \citep{Freitas:1983a}.  This theory still imposes restrictions on the nature of ETI, in that while it might accommodate a few unnoticed SRPs, it is unclear how it would fare against the potential for millions of SRPs (as described in section \ref{label_section_EstimatingtheSRPPopulation}).  Likewise, this theory strictly excludes pervasive colonization since we can all agree no such effort has reached us.  Nevertheless, it is compelling and has implications for future SETI programs: we should search more aggressively within our solar system.  Freitas has also described what to look for and where to look for it \citep{Freitas:1983b}.

Furthermore, few of the arguments on either side of the debate have much implication for other galaxies.  One attempt at such an argument is by Sagan and Newman \citep{SaganNewman:1983}, in which they conclude -- with what can only be regarded as scathing sarcasm -- that the notion of SRPs requires  every galaxy in the cosmos to be completely mass-converted into probes.  Therefore, they conclude that SRPs are a flawed theory.  As explained in sections  \ref{label_section_SelfReplicatingProbeVoluntaryRefrainment} and \ref{label_section_BiologicalBias}, we believe the more likely explanation is that their glib corollary speaks first to an overestimation of the probability of a successful cancerous mutation, and second to a deep biological bias with regard to computerized intelligence.  If we disregard Sagan and Newman's claim, and accept a subluminal speed limit, then it is perfectly reasonable for extragalactic ETI to exist.

This conclusion offers plenty of headroom in the argument over who should be viewed as an optimist as opposed to a pessimist.  Those who argue against numerous intragalactic ETI are often labeled pessimists.  However, if extragalactic ETI are brought back into the fold, and given that there are several 100 billion galaxies in the universe, we can easily permit comparable numbers of ETI without the slightest conflict with the Fermi Paradox, clearly not a pessimistic perspective.  Of greater import, this idea has tremendous implications for SETI: we need to search other galaxies.

\section{SETI}
\label{label_section_SETI}

One way of categorizing potential SETI targets is to break them into three distances ranges: the solar system, the galaxy, and the universe.  Most SETI programs have focused on the galaxy.  Ironically, of the three potential targets, we may be directing most of our efforts on the least likely category to yield fruit.  Ideally, our society would value research of this sort to such a degree that no hard choices would have to be made and we could simply search everywhere.  However, if choices must be made, perhaps we should target at least a few SETI programs to the solar system to look for SRPs and to other galaxies to look for possible indicators of intelligence.  As a starting point, we would recommend considering nearby face-on spiral galaxies residing well above the galactic plane of the Milky Way, \ie galaxies to which we ourselves appear face-on.  Thus, EM transmissions will be the least extincted by cosmic dust in either galaxy, thereby maximizing the chance of detection.  Possible candidates might include M31 and M33 (whose oblique inclinations and low galactic latitudes are compensated for by their remarkable proximity), NGC300, NGC2403, M51, M81, M101, and many others.

\section{Conclusion}
\label{label_section_Conclusion}

This paper was arranged in three parts.  First, we introduced SRPs, presented the prevalent arguments against them, and showed that such arguments leave room for future SRP consideration.  Namely, we proposed that recent literature has been overzealous in its exclusion of SRPs and we encourage their return to the field.

Second, we presented percolation theory and its nonsociological explanation for the Fermi Paradox.  We then showed that the theory can be extended in very reasonable ways which undermine its primary conclusion that galactic expansion might be intrinsically bounded.

Third, we reviewed two theories of interstellar societal collapse and showed a few counter-arguments to each theory.  Furthermore, we introduced ITB theory and showed that its implications might suggest a fundamental error in such theories.

We then discussed one additional paper theorizing that interstellar societies shrink back to their homeworlds and explained that the model involves a number of unlikely assumptions.  Following this final analysis, we described the best theory yet offered on the Fermi Paradox which permits intragalactic ETI, namely that exploration probes may currently reside in our solar system, yet undiscovered.  Lastly, we offered our thoughts on how to design future SETI programs so as to maximize the likelihood of success.

\section{Acknowledgements}
\label{label_section_Acknowledgements}

We wish to thank Angeline Madrid for editing and our Facebook/Google+ intellectual inner circle -- James Horey, Aaron Clauset, Kshanti Greene, Terran Lane, Diane Oyen, and Marlow Weston -- for helping to name the ITB theory and refine the mathematical formulations in section \ref{label_section_RefiningtheDrakeEquation}.


\bibliographystyle{model2-names}
\bibliography{bibliography}

\begin{thebibliography}{24}
\expandafter\ifx\csname natexlab\endcsname\relax\def\natexlab#1{#1}\fi
\expandafter\ifx\csname url\endcsname\relax
  \def\url#1{\texttt{#1}}\fi
\expandafter\ifx\csname urlprefix\endcsname\relax\def\urlprefix{URL }\fi
\providecommand{\eprint}[2][]{\url{#2}}
\providecommand{\bibinfo}[2]{#2}
\ifx\xfnm\relax \def\xfnm[#1]{\unskip,\space#1}\fi
\bibitem[{Bezsudnov and Snarskii(2010)}]{BezsudnovSnarskii:2010}
\bibinfo{author}{Bezsudnov, I.}, \bibinfo{author}{Snarskii, A.},
  \bibinfo{year}{2010}.
\newblock \bibinfo{title}{Where is everybody? -- wait a moment ... new approach
  to the fermi paradox}.
\newblock \bibinfo{journal}{ArXiv e-prints .} \eprint{1007.2774v1}.
\bibitem[{Bjoerk(2007)}]{Bjoerk:2007}
\bibinfo{author}{Bjoerk, R.}, \bibinfo{year}{2007}.
\newblock \bibinfo{title}{Exploring the galaxy using space probes}.
\newblock \bibinfo{journal}{International Journal of Astrobiology} .
\bibitem[{Brin(1983)}]{Brin:1982}
\bibinfo{author}{Brin, G.D.}, \bibinfo{year}{1983}.
\newblock \bibinfo{title}{The 'great silence': the controversy concerning
  extraterrestrial intelligent life}.
\newblock \bibinfo{journal}{Royal Astronomical Society Quarterly Journal}
  \bibinfo{volume}{24}, \bibinfo{pages}{283--309}.
\bibitem[{Chyba and Hand(2005)}]{ChybaHand:2005}
\bibinfo{author}{Chyba, C.F.}, \bibinfo{author}{Hand, K.P.},
  \bibinfo{year}{2005}.
\newblock \bibinfo{title}{Astrobiology: The study of the living universe}.
\newblock \bibinfo{journal}{Annual Review of Astronomy and Astrophysics}
  \bibinfo{volume}{43}, \bibinfo{pages}{31--74}.
\bibitem[{Cotta and Morales(2009)}]{CottaMorales:2010}
\bibinfo{author}{Cotta, C.}, \bibinfo{author}{Morales, A.},
  \bibinfo{year}{2009}.
\newblock \bibinfo{title}{A computational analysis of galactic exploration with
  space probes: Implications for the fermi paradox}.
\newblock \bibinfo{journal}{Journal of the British Interplanetary Society}
  \bibinfo{volume}{62}, \bibinfo{pages}{82--88}.
\bibitem[{Diamond(2005)}]{Diamond:2005}
\bibinfo{author}{Diamond, J.M.}, \bibinfo{year}{2005}.
\newblock \bibinfo{title}{Collapse: How Societies Choose to Fail Or Succeed}.
\newblock \bibinfo{publisher}{Penguin Books}.
\bibitem[{Dyson(1968)}]{Dyson:1968}
\bibinfo{author}{Dyson, F.J.}, \bibinfo{year}{1968}.
\newblock \bibinfo{title}{Interstellar transport}.
\newblock \bibinfo{journal}{Physics Today} , \bibinfo{pages}{41--45}.
\bibitem[{Forgan(2009)}]{Forgan:2009}
\bibinfo{author}{Forgan, D.H.}, \bibinfo{year}{2009}.
\newblock \bibinfo{title}{A numerical testbed for hypotheses of
  extraterrestrial life and intelligence}.
\newblock \bibinfo{journal}{International Journal of Astrobiology}
  \bibinfo{volume}{8}, \bibinfo{pages}{121--131}.
\bibitem[{Gregersen(2009)}]{Gregersen:2009}
\bibinfo{editor}{Gregersen, E.} (Ed.), \bibinfo{year}{2009}.
\newblock \bibinfo{title}{The Milky Way and Beyond: Stars, Nebulae, and Other
  Galaxies}.
\newblock \bibinfo{publisher}{The Rosen Publishing Group}.
\bibitem[{Group(1978)}]{Daedalus:1978}
\bibinfo{author}{Group, P.D.S.}, \bibinfo{year}{1978}.
\newblock \bibinfo{title}{Project Daedalus : the final report on the BIS
  starship study}.
\newblock \bibinfo{publisher}{Journal of the British Interplanetary Society}.
\bibitem[{Haqq-Misra and Baum(2009)}]{MisraBaum:2009}
\bibinfo{author}{Haqq-Misra, J.}, \bibinfo{author}{Baum, S.D.},
  \bibinfo{year}{2009}.
\newblock \bibinfo{title}{The sustainability solution to the fermi paradox}.
\newblock \bibinfo{journal}{Journal of the British Interplanetary Society}
  \bibinfo{volume}{62}, \bibinfo{pages}{47--51}.
\bibitem[{Hart(1975)}]{Hart:1975}
\bibinfo{author}{Hart, M.H.}, \bibinfo{year}{1975}.
\newblock \bibinfo{title}{An explanation for the absence of extraterrestrials
  on earth}.
\newblock \bibinfo{journal}{Q.J. R. Astr. Soc.} \bibinfo{volume}{16},
  \bibinfo{pages}{128--135}.
\bibitem[{Jones(1981)}]{Jones:1981}
\bibinfo{author}{Jones, E.M.}, \bibinfo{year}{1981}.
\newblock \bibinfo{title}{Discrete calculations of interstellar migration and
  settlement}.
\newblock \bibinfo{journal}{Icarus} \bibinfo{volume}{46},
  \bibinfo{pages}{328--336}.
\bibitem[{Jr.(1983a)}]{Freitas:1983a}
\bibinfo{author}{Jr., R.A.F.}, \bibinfo{year}{1983}a.
\newblock \bibinfo{title}{Extraterrestrial intelligence in the solar system:
  Resolving the fermi paradox}.
\newblock \bibinfo{journal}{J. Brit. Interplanet. Soc.} \bibinfo{volume}{36},
  \bibinfo{pages}{496--500}.
\bibitem[{Jr.(1983b)}]{Freitas:1983b}
\bibinfo{author}{Jr., R.A.F.}, \bibinfo{year}{1983}b.
\newblock \bibinfo{title}{If they are here, where are they? observational and
  search consideration}.
\newblock \bibinfo{journal}{Icarus} \bibinfo{volume}{55}.
\bibitem[{Landis(1998)}]{Landis:1998}
\bibinfo{author}{Landis, G.A.}, \bibinfo{year}{1998}.
\newblock \bibinfo{title}{The fermi paradox: An approach based on percolation
  theory}.
\newblock \bibinfo{journal}{J. Brit. Interplanet. Soc.} \bibinfo{volume}{51},
  \bibinfo{pages}{163--166}.
\bibitem[{Matloff(2005)}]{Matloff:2005}
\bibinfo{author}{Matloff, G.L.}, \bibinfo{year}{2005}.
\newblock \bibinfo{title}{Deep Space Probes: To the Outer Solar System and
  Beyond}.
\newblock \bibinfo{publisher}{Praxis Publishing Ltd.}
\bibitem[{McInnes(2002)}]{McInnes:2002}
\bibinfo{author}{McInnes, C.}, \bibinfo{year}{2002}.
\newblock \bibinfo{title}{The light cage limit to interstellar expansion}.
\newblock \bibinfo{journal}{Journal of the British Interplanetary Society}
  \bibinfo{volume}{55}, \bibinfo{pages}{279--284}.
\bibitem[{Sagan and Newman(1983)}]{SaganNewman:1983}
\bibinfo{author}{Sagan, C.}, \bibinfo{author}{Newman, W.},
  \bibinfo{year}{1983}.
\newblock \bibinfo{title}{The solipsist approach to extraterrestrial
  intelligence}.
\newblock \bibinfo{journal}{Quarterly Journal of the Royal Astronomical
  Society} \bibinfo{volume}{24}, \bibinfo{pages}{113--121}.
\bibitem[{Shostak(2010)}]{Shostak:2010}
\bibinfo{author}{Shostak, S.}, \bibinfo{year}{2010}.
\newblock \bibinfo{title}{What et will look like and why should we care}.
\newblock \bibinfo{journal}{Acta Astronautica} \bibinfo{volume}{67},
  \bibinfo{pages}{1025--1029}.
\bibitem[{Tipler(1980)}]{Tipler:1980}
\bibinfo{author}{Tipler, F.J.}, \bibinfo{year}{1980}.
\newblock \bibinfo{title}{Extraterrestrial intelligent beings do not exist}.
\newblock \bibinfo{journal}{Quarterly Journal of the Royal Astronomical
  Society} \bibinfo{volume}{21}, \bibinfo{pages}{267--281}.
\bibitem[{Turner(1978)}]{Turner:1978}
\bibinfo{author}{Turner, J.M.}, \bibinfo{year}{1978}.
\newblock \bibinfo{title}{Fetus into Man}.
\newblock \bibinfo{publisher}{Harvard University Press}.
\bibitem[{Walters et~al.(1980)Walters, Hoover and Kotra}]{Walters:1980}
\bibinfo{author}{Walters, C.}, \bibinfo{author}{Hoover, R.},
  \bibinfo{author}{Kotra, R.K.}, \bibinfo{year}{1980}.
\newblock \bibinfo{title}{Interstellar colonization: A new parameter for the
  drake equation}.
\newblock \bibinfo{journal}{Icarus} \bibinfo{volume}{41},
  \bibinfo{pages}{193--197}.
\bibitem[{Wiley(2011)}]{Wiley:2011}
\bibinfo{author}{Wiley, K.B.}, \bibinfo{year}{2011}.
\newblock \bibinfo{title}{Implications of computerized intelligence on
  interstellar travel}.
\newblock \bibinfo{journal}{H+ Magazine} .

\end{thebibliography}

\end{document}